\documentclass{elsart}

\usepackage{amssymb}
\usepackage{graphicx}

\begin{document}

\begin{frontmatter}



\title{Dynamics of rotating Bose-Einstein condensates probed by Bragg scattering}


\author{S. R. Muniz}
\ead{sergio.muniz@physics.gatech.edu}
\author{D. S. Naik, M.  Bhattacharya}
\author{C. Raman}
\ead{craman@gatech.edu}

\address{School of Physics, Georgia Institute of Technology, Atlanta, Georgia, USA 30332-0430}

\begin{abstract}
Gaseous Bose-Einstein condensates (BECs) have become an important test bed for studying the dynamics of quantized vortices.  In this work we use two-photon Doppler sensitive Bragg scattering to study the rotation of sodium BECs.  We analyze the microscopic flow field and present laboratory measurements of the coarse-grained velocity profile.  Unlike time-of-flight imaging, Bragg scattering is sensitive to the direction of rotation and therefore to the phase of the condensate.  In addition, we have non-destructively probed the vortex flow field using a sequence of two Bragg pulses.

\end{abstract}

\end{frontmatter}

\section{Introduction}

Vortices are a hallmark of superfluids and have applications throughout the study of fluid mechanics and condensed matter physics \cite{donn91,bare01,nels01}.  Recently, gaseous Bose-Einstein condensates (BECs) under rotation have become an important test bed for predictions of the behavior of quantized vortices, which form highly regular lattices \cite{madi00,abos01latt,halj01,hodb01}.  In these gases, time-of-flight (TOF) imaging has been an indispensable tool for observing the rotation of the cloud through the detection of the vortex cores.  However, this technique only measures the superfluid density, and not the phase of the macroscopic wavefunction.  Phase measurements would provide a more complete picture of vortex states, especially useful in cases where individual vortices might not be easily detected.  In atomic gases, one can apply powerful optical and spectroscopic techniques \cite{kozu99bragg,sten99brag,sims00,thei04,katz04} that afford new possibilities for performing phase measurements.

In this work, we explore two photon Bragg scattering as a tool that provides information about the phase of a rotating Bose-Einstein condensate.  We focus on both the microscopic and macroscopic signatures of the vortices in section 2 of the paper, and present experimental data using this technique in section 3.

\section{Vortex Lattices and the Bragg Method}

\subsection{Velocity Field of Rotating Condensates}

Vortex states in BECs have gained interest in recent years.  Here we will give a brief introduction to their properties, and refer the reader to useful review articles by Fetter and Kevrekidis for further details \cite{fett01review,kevr04}.  Our starting point is a gas of bosonic atoms for which the order parameter $\Psi(\vec{x},t)$ satisfies the Gross-Pitaevskii equation \cite{pita03book}
\begin{center}
\begin{equation}
i \hbar \frac{\partial \Psi}{\partial t} = \left(-\frac{\hbar^2}{2 M} \nabla^2 + V_{ext}(\vec{x},t) + g |\Psi|^2\right ) \Psi \nonumber
\end{equation}
\end{center}
where $\hbar = \frac{h}{2 \pi}$ is the reduced Planck constant, $M$ is the atomic mass and $V_{ext}$ is the sum of all external potentials including the trapping potential.  $\Psi$ is a single-particle Schrodinger wavefunction for which $\int |\Psi|^2 d^3 x = N$, the total atom number.  It approximates the many body physical description in the limit of weak interactions between the atoms and very low temperatures, both of which can be satisfied in the laboratory for a wide range of conditions.  Since $\Psi$ is in general complex, it may be written as $\sqrt{n(\vec{x},t)} e^{i S(\vec{x},t)}$, where $n$ is the superfluid density and $S$ the phase of the wavefunction.  The superfluid velocity field $\vec{v} = \frac{\hbar}{M} \nabla S$ is proportional to the gradient of the phase.  Since phase is only defined modulo $2 \pi$, it is readily apparent that the circulation $\oint \vec{v} \cdot \vec{dl} = m \times \frac{h}{M}$ must be quantized, where $m = 1,2,3....$.  The simplest state for which this condition is satisfied is a single vortex state with $m=1$.

For our purposes, we will assume that we have a harmonically trapped Bose-Einstein condensate in the oblate potential of a ``TOP'', or time-orbiting potential magnetic trap.  This trap has azimuthal symmetry in the $x-y$ plane.  It consists of a rapidly rotating magnetic bias field superimposed on a static quadrupole magnetic trap.  The net effect is to move the magnetic field zero outside of the cloud, thus preventing nonadiabatic spin flips \cite{petr95}, and resulting in a time-averaged potential $V_{ext} = \frac{1}{2} M \left(\omega_t^2(x^2+y^2)+ \omega_z^2 z^2 \right)$.  Within this potential, a singly quantized vortex line aligned with the $z$ direction has a phase $S = \phi$, where $\phi = $tan$^{-1} (y/x)$ is the azimuthal angle.  Thus the velocity field
\begin{equation}
    \vec{v}_1 (x,y) = \frac{\hbar}{M} \nabla S = \frac{\hbar}{M} \frac{\hat{y}x -\hat{x}y}{x^2+y^2}
\label{eqn:single_vortex}\end{equation}
where $\hbar$ is the Planck constant and $M$ is the mass of the atom.

\begin{figure}[tbph]
\begin{center}
\includegraphics[width = \textwidth]{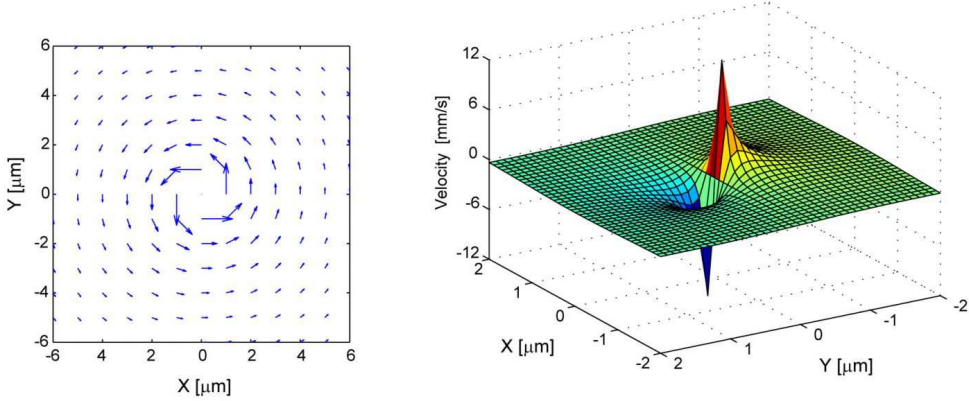}
\end{center}
\caption{Microscopic velocity field of a vortex state.  Velocity field of a single vortex is shown on the left.  The arrows indicate the direction of the local velocity vector $\vec{v}_1 (x,y)$ whose magnitude is proportional to the arrow length.  The projection $v_{1x}$ as measured by Bragg scattering is plotted on the right.}
\label{single_vortex}
\end{figure}

In Figure \ref{single_vortex}, we show the vector velocity field of a single vortex from Eqn.\ \ref{eqn:single_vortex}, as well as its $\hat{x}$-projection $\vec{v}_{1x} = \vec{v}_1 \cdot \hat{x}$ as a 3-dimensional surface.  From Eqn. \ref{eqn:single_vortex} and the plots, one can see that although the vector magnitude $|\vec{v_1}|$ has azimuthal symmetry, the projection $\vec{v}_{1x}$ clearly does not --  the x-velocity is largest along the $y$-axis and goes to zero along the $x$-axis.  The region $\rho = \sqrt{x^2+y^2}< \xi$ that contains the singularity at the origin has been removed from the plots for clarity.

\begin{figure}[tbph]
\begin{center}
\includegraphics[width = 0.75\textwidth]{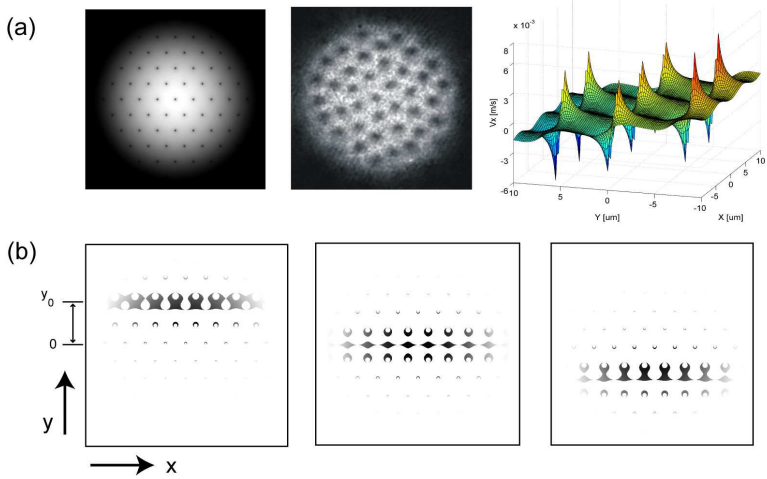}
\end{center}
\caption{Microscopic velocity field of a vortex lattice.  Shown are (a) column density profile in the trap (approximate solution, left) and TOF images (experiment, center).  On the right, the velocity projection $v_{Nx}$ is shown for a vortex lattice according to Eqn. \ref{eqn:many_vortex}.  (b) Spectral density of Bragg scattered atoms for $-2.7\pm0.3$ mm/s (left), $\pm0.3$ mm/s (center) and $2.1\pm0.3$ mm/s (right), showing microscopic as well as macroscopic flow.}
\label{Vortex-Lattice}
\end{figure}
For large angular momentum, the superfluid typically breaks up into a lattice of $N_v$ singly-quantized vortices that form a triangular  structure \cite{abos01latt}.  We have shown the corresponding column density profile of the atoms, proportional to $\int |\Psi|^2 dz $, in Figure \ref{Vortex-Lattice}a.  In the vicinity of each vortex core, the particle density $\rightarrow 0$ within a region of radius $\approx \xi$, where $\xi$ is the ``healing length'', the microscopic length scale that emerges from the GPE.  Except at the condensate boundaries, the overall density distribution of the trapped gas is well approximated by the Thomas-Fermi profile $n(\rho,z) = |\Psi_0|^2 = 1-z^2/R_z^2-\rho^2/R_{TF}^2$.  The Thomas-Fermi radii are $R_z$ and $R_{TF}$ along the $z$ and $\rho$ direction, respectively \cite{pita03book}.  Typically, $\xi = 0.5 \mu$m, $R_{TF} = 37 \mu$m and $R_z = 13\mu$m for our experimental parameters.  To generate the column density profile, we multiply the Thomas-Fermi solution $\Psi_0$ by a variational function $f(x,y)$ for the vortex density.  We used $f(x,y) = \prod_{i=1}^{N_v} \frac{s_i}{\sqrt{2+s_i^2}}$, where $s_i = \sqrt{(x-x_i)^2+(y-y_i)^2}/\xi$ and the $N_v$ vortices are located at $\vec{\rho_i} = (x_i,y_i)$, with $i = 1,...N_v$ \cite{peth02book}.  This allows the density to vanish at each singularity and gives us an approximate solution to the Gross-Pitaevskii equation which is valid so long as the vortex separation $d$ greatly exceeds the core size $\xi$.  For our parameters $d \simeq R_{TF}/\sqrt{N_v/\pi} = 10 \mu$m $\approx 20 \times \xi$.  Adjacent to the vortex distribution, we have also shown an experimental image of the column density profile which we have observed in a {\em time-of-flight} image of a rotating BEC, which confirms the lattice structure of the vortices.  This method involves turning off the trap at time $t_0$ and imaging the column density after a variable time of free flight, and the image roughly corresponds to the velocity distribution of the atoms just before $t_0$.

The corresponding velocity field of a rapidly rotating BEC can be approximated as the sum over all vortices.  This is a valid approximation in the limit where the vortex separation greatly exceeds the core size:
\begin{equation}
    \vec{v}_N (x,y) = \frac{\hbar}{M} \sum_{i=1}^{N_v} \frac{\hat{y}(x-x_i) -\hat{x}(y-y_i)}{(x-x_i)^2+(y-y_i)^2}
\label{eqn:many_vortex}\end{equation}
In Figure \ref{Vortex-Lattice}a (rightmost graph), we have plotted $\vec{v}_N \cdot \hat{x}$ for a hexagonal lattice containing several vortices.  One can clearly observe that the background velocity (i.e. away from the singularities) increases roughly linearly with the $y$-coordinate.  That is, when one coarse-grains over the velocity fields of the individual vortices, the resulting field is that of a rigid body rotation $v_x = \Omega y$, where $\Omega$ is the rate of rotation of the lattice \cite{nozi99}.  There are also intriguing signatures of the core regions in Figure \ref{Vortex-Lattice}b, which we will explore below.

\subsection{Bragg Scattering from Rotating Condensates}

\subsubsection{Technique}
Time-of-flight images such as those we have shown in Figure \ref{Vortex-Lattice}a typically measure only the column density distribution of the atoms.  That is, the signal in the image is proportional to $|\Psi|^2$, and does not directly measure the phase $S$ of the wavefunction\footnote{The amplitude of $\Psi$ does depend on the phase $S$, however, since the two are coupled during the time-of-flight evolution.}.  One technique for probing $S$ is two photon Bragg scattering \cite{sten99brag,sims00}, which is sensitive to the velocity of the atoms, which is $\propto \nabla S$.  The full details of Bragg scattering process are beyond the scope of this paper, and therefore, we will concentrate only on a few key features which are relevant to our data. More details on Bragg scattering can be found elsewhere \cite{stam01bragrev,blak02}.

\begin{figure}[tbph]
\begin{center}
\includegraphics[width = \textwidth]{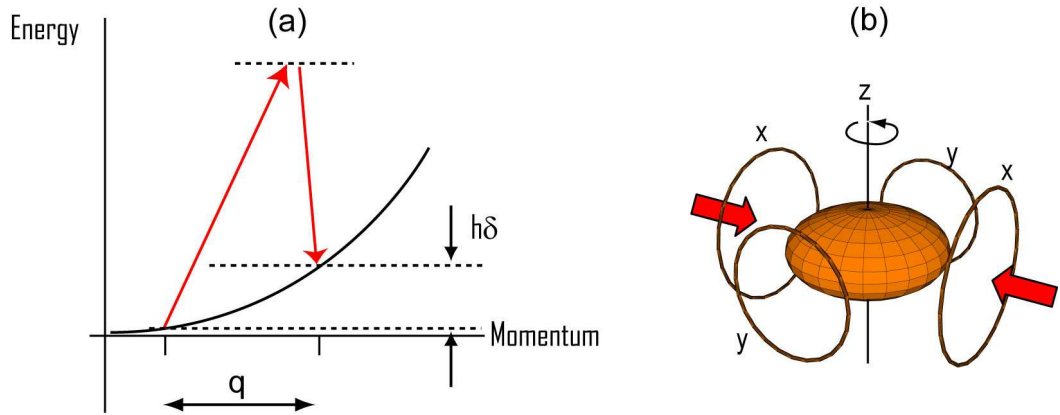}
\end{center}
\caption{(a) Energy-momentum relation for atoms in a BEC showing the momentum transfer $q$ from the two-photon process.  The Bragg resonance occurs at a frequency difference $\delta$ between the two laser beams that satisfies both energy and momentum conservation.  (b) Diagram of experimental geometry.  Vortices are created by rotating a BEC about the $z$-axis by phase control of the transverse fields produced in coil pairs $x$ and $y$ that control the TOP (time-orbiting potential) trap.  The Bragg beam containing frequencies $\omega$ and $\omega+2 \pi \delta$ is applied along the $x$-direction and retroreflected.}
\label{fig:bragg}
\end{figure}

In brief, the Bragg method employs two laser beams with frequencies $\omega$ and $\omega+2 \pi \delta$.  An atom scatters a photon from one laser beam into another.  The net result is to impart to it a momentum $\textbf{q}$.  For counterpropagating laser beams of wavelength $\lambda$, $q=|\textbf{q}| = 2h / \lambda$ is twice the momentum of a single photon.  The two photon process and the energy-momentum relation are shown schematically in Figure \ref{fig:bragg}a.  For a condensate, $q/M$ is typically  much greater than the initial velocity of the atoms, and therefore, the diffracted cloud can be easily distinguished from the non-diffracted atoms.  This is because the former have traveled an additional distance $\approx q/M \times t_{tof}$ during the time-of-flight $t_{tof}$ after the trap has been shut off.  The two clouds can then be separated in the images, as we show below.  There is a resonance in the scattering of light when
\begin{equation}
\delta = \delta_{MF}+ \frac{q^2}{2 M h} + \frac{\textbf{q} \cdot \textbf{v}}{h}
\label{eq:delta}
\end{equation}
which expresses the conservation of momentum and energy.  In the above equation, the second term is the recoil energy, which must be provided by the energy difference between the two photons.  For sodium atoms near the principal resonance, $\frac{q^2}{2 m h}=100$ kHz.  The third term is simply the Doppler shift, which makes the Bragg technique velocity sensitive.  It is this term which is of primary importance to this work, as the Bragg process selects a group of atoms with the same projection of velocity $v_x$ along the direction of the momentum transfer $\textbf{q} = q \hat{x}$.  For a trapped BEC, one also has to consider the effects of interactions.  In the mean-field and local density approximations \cite{blak02,zamb00}, this causes an extra $\delta_{MF} = \frac{4 \mu}{7 h}$ frequency shift\footnote{The inhomogeneous density distribution also causes a broadening of the resonance.  For our parameters, this is discussed in \cite{muni06}.}, where $\mu$ is the chemical potential.  This shift is a consequence of the Bogoliubov dispersion relation for condensate excitations, as discussed in reference \cite{sten99brag}.  Free particles have $\delta_{MF} = 0$.  In our case, for a stationary condensate, the Bragg resonance is peaked at a frequency $\delta_o = \frac{q^2}{2 m h} + \delta_{MF} \approx 101$ kHz.  Thus  $\delta = \delta_o + \frac{\textbf{q} \cdot \textbf{v}}{h}$.

\subsubsection{Microscopic Flow Field}
Spatially selective Bragg scattering in principle allows one to ``peer into'' the complex velocity flow field of Eqn. \ref{eqn:many_vortex}.  One may understand this in an intuitive way by considering the velocity surface $v_x(x,y)$ in Figure \ref{Vortex-Lattice}. For a given Bragg beam detuning $\delta$ we select a velocity $v_{x,0}$ through Eqn.\ \ref{eq:delta}.  The region of space containing atoms that are moving at that velocity is given by the solution to the equation $v_x(x,y) = v_{x,0}$.  Geometrically, this represents the intersection of the surface with a plane at height $v_{x,0}$, and is simply a velocity contour map.  In Figure \ref{Vortex-Lattice}b, we show the resulting {\em spatial distribution} of atoms that are Bragg diffracted.  The grey scale is proportional to the number of atoms within a velocity range of $\pm0.3$mm/s, typical of the experimental resolution.  Two features are clearly visible.  For one, there is a broad horizontal band of atoms that are resonant with the laser whose center of mass y-coordinate $y_0$ varies as one varies the detuning of the laser beams.  $y_0$ satisfies the rigid body condition $v_x = \Omega y$.  We have varied the detuning to map out $y_0$ as a function of the velocity experimentally \cite{muni06}.

In addition, the spatial distributions contain information about the {\em microscopic} flow field of the vortices.  The bands at constant $y_0$ in Figure \ref{Vortex-Lattice}b are not uniform; rather, there is a local structure to the velocity field associated with the flow around individual vortices.  This leads to beautiful patterns in the spatial profile of the diffracted atoms, including crescent and diamond shapes.  These patterns are {\em imprinted} by the Bragg beams through the velocity selection.  The existence of spatial signatures of the vortex state was first discovered by Blakie et al.\ \cite{blak01}, who studied the Bragg scattering from a {\em single} vortex.  They used a mean-field approach, incorporating the laser beam potential into the Gross-Pitaevskii equation to derive the spatial and temporal evolution of the Bragg scattered atoms.  Observing signatures of these patterns in time-of-flight imaging is complicated by the mean-field evolution during the time-of-flight expansion, as discussed in \cite{muni06}.  In the following section we also discuss the necessary tradeoff between spatial and velocity resolution.

\subsubsection{Spectral Width}
The spectral width of the Bragg resonance is of critical importance, and is discussed in detail in a number of works.  The finite temporal width $\tau_{B}$ of the Bragg pulse creates spectral broadening.  Through the Doppler shift, this limits the velocity resolution $\Delta v_x$ to
\begin{equation}
\Delta v_x = \frac{1}{2 \pi \tau_B}\frac{h}{q }
\end{equation}

A rotating BEC also has a spatial distribution of velocities.  In order to achieve a spatial resolution $\Delta x$ along the direction of the Bragg beams, the pulse duration must be short enough that the recoiling atoms from regions of the condensate that are separated by $\Delta x$ do not spatially overlap during the pulse.  Therefore, \begin{equation}
\Delta x = \frac{q}{M} \tau_B
\end{equation}

In our experiments we choose a time $\tau_B \approx 250 \mu$s.  The recoil velocity is $q/M = 5.9$ cm/s, while typical rotational velocities are $v_{rot} \approx 3.5$ mm/s and the Thomas-Fermi diameter is $D_{TF} = 74 \mu$m.  This yields $\Delta v_x \approx v_{rot}/20$ and $\Delta x \approx D_{TF}/5$, which is a balance between good velocity and spatial resolution to observe the overall structure of the rotation.

It is interesting to note the trade-off between high spatial and velocity resolutions, since their product $\Delta x \Delta v_x = \hbar/M$ is a constant. For very short pulses $\approx \xi M/q = 8 \mu$s for our system, the vortex core regions could be resolved optically according to the above criterion; however, there will be a substantial spread of the velocities of the atoms which are Bragg diffracted.

\section{Experiment and Data}

In our experiments we produce a sodium BEC with typically $1-3\times 10^{6}$ atoms in a time-averaged orbiting potential (TOP) trap \cite{petr95}, according to the method described in \cite{muni06}.  Parameters for the trap are a radial gradient  $B_\rho' = 12 $ Gauss/cm and a bias rotation of $\omega _{TOP}=2\pi\times5$ kHz. The measured transverse oscillation frequency is $\omega_{\rho}=2\pi \times 31$ Hz, with $\omega_z =\sqrt{8} \omega_{\rho}$.  For details of experimental techniques, the reader is referred to several excellent review articles \cite{cornkett99var}.

We produce the vortex-lattice by creating a rotating elliptical asymmetry in the horizontal \textit{x-y} plane of the TOP trap \cite{hodb01}.  The TOP trap employs a fast rotating bias field $\vec{B} = (B_x(t),B_y(t))$ at a frequency $\omega_{TOP}$ much greater than that of the atomic motion.  This creates a time-averaged harmonic potential.  Therefore, we can create a slowly rotating elliptical potential by superimposing slow variations upon the fast oscillation of the bias field.  To produce the fields, we combine the signals of two digital frequency synthesizers operating at frequencies $\omega_{1} = \omega_{TOP}+ \omega_{AR}$ and $\omega_{2} = \omega_{TOP}- \omega_{AR}$.  These signals are each split, phase shifted and summed together to produce the two fields:  $B_x(t) = B_0 \cos(\omega_1 t)+\epsilon \cos(\omega_2 t)$ and $B_y(t) = B_0 \sin(\omega_1 t)-\epsilon \sin(\omega_2 t)$, where $\epsilon$ and $\omega_{AR}$ are the amplitude and frequency of the rotating asymmetry, respectively.  The two currents are individually amplified using 100 Watt car audio amplifiers, and capacitively coupled to a pair of Helmholtz coils of approximately 10 cm diameter along the $x$ and $y$ directions, respectively (see Figure \ref{fig:bragg}b).  In order to maximize the number of vortices, we chose $\omega_{AR}= 2\pi \times 22$ Hz, which is very close to the frequency $\simeq 0.7 \omega_{\rho}$ that drives the quadrupole mode in our harmonic trap \cite{madi01}. After applying the rotating asymmetry for 1.5 seconds, it was turned off and the atomic cloud allowed to equilibrate in the trap for another 1 to 1.5 seconds. This procedure reliably created vortex lattices with approximately 40 $\pm$ 10 vortices, as shown in Fig. 2a.
\begin{figure}[tbph]
\begin{center}
\includegraphics[width = \textwidth]{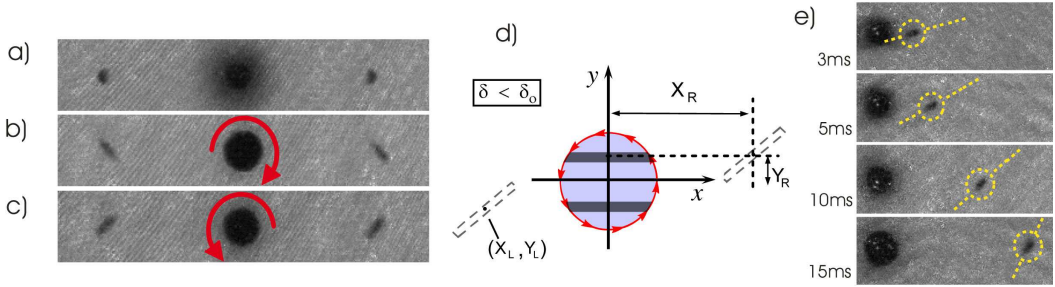}
\end{center}
\caption{Vortices probed by Bragg scattering.  Outcoupled atoms (to the far right and left within each image) showed no particular structure for non-rotating clouds (a), whereas from vortices (b) and (c), the outcoupled atoms were tilted according to the direction of rotation. Images (a-c) were taken at 10 ms TOF.  The tilt angle increases with respect to time of flight, as shown in (e). Each pair of Bragg frequencies is resonant with a thin strip of atoms parallel to the $x$-axis, as illustrated in (d). All images were taken at $\delta = 102$kHz.}
\label{Figure4: Spatial-Structures}
\end{figure}

\subsection{Spatially Resolving the Velocity Field}

After producing vortices and allowing the lattice to equilibrate, we pulse the Bragg diffracting beams along the $x$-direction (see Figure \ref{fig:bragg})b) for a time $\tau_B$, while the atoms are still in the trap. The Bragg beams are detuned by 1.7 GHz from the $F=1$ to $F'=2$ resonance, and are created by back-reflecting a single beam that contains two frequencies $\omega_{L}$ and $\omega_{L} +  2\pi \delta$.  $\delta$ is the difference between the frequencies of two rf synthesizers that are used to drive a single acousto-optic modulator.  This creates 2 groups of diffracted atoms propagating to the left and to the right, respectively.  We applied a Bragg pulse of square shape with $\tau_B=250 \mu s$, and then turned off the magnetic trap within $100 \mu$s.  The atoms expanded for a variable $t_{tof}$ before we took an absorption image using laser light resonant with the $F=1\rightarrow2$ transition in a $250\mu$s pulse.  The result is shown in Figure \ref{Figure4: Spatial-Structures}.  We can clearly observe {\em spatial structures} in the outcoupled atom cloud arising from the rotation of the cloud.  In figure \ref{Figure4: Spatial-Structures}a one can see the diffraction from an initially stationary condensate, and the diffracted (outcoupled) atoms appear to the right and left of the stationary condensate.  No particular structure is visible.  However, in Figure \ref{Figure4: Spatial-Structures}b, we have initially prepared a vortex lattice, which causes the diffracted atoms to form a tilted, elongated spatial pattern.  Moreover, when we reversed the direction of the applied rotation (by replacing $\omega_{AR}\rightarrow -\omega_{AR}$), the tilt angle with respect to the $y$-direction reverses, as shown in Figure \ref{Figure4: Spatial-Structures}c.

We can understand our observations in terms of the coarse-grained velocity field discussed earlier:  $v = \vec{\Omega} \times \vec{r}$, with $\Omega = |\vec{\Omega}|$ proportional to the number of vortices.  Since the Bragg process selects a group of atoms with the same $v_x = \Omega y$, the resonance condition is given by $\delta = \delta_o + 2y\Omega / \lambda$. Therefore, for a spectrally narrow Bragg pulse, with $\delta <\delta_0$ and a counter-clockwise rotation, the resonance corresponds to a thin, horizontal band of atoms with $y>0$ for atoms which are Bragg scattered to the right, and $y<0$ for atoms scattered to the left (the dark shaded regions in Figure \ref{Figure4: Spatial-Structures}d). This band is identical to what is shown in Figure \ref{Figure4: Spatial-Structures}c, and within it there is a detailed microscopic structure near the vortex cores which is not resolved in our current experiment.  As this band of atoms moves, the spread in velocities in the $x-y$ plane causes part of the band to move up while another part moves down.  Thus it forms a tilted stripe whose angle increases with time, as observed in Figure \ref{Figure4: Spatial-Structures}e. At long TOF the stripe should become fully stretched along the vertical axis of the images.

A calculation of the rotation frequency from the evolution of the tilt angle as function of time (see Fig. \ref{Figure4: Spatial-Structures}e), $\theta(t)=\arctan( \Omega t)$, results in $\Omega=2\pi \times( 15.4 \pm 1.1$ Hz).  Moreover, one can use the locations of the diffracted cloud ($X_R,X_L$ and $Y_R,Y_L$ in Figure \ref{Figure4: Spatial-Structures}d) to spatially map the velocity field, which results in a similar value for the rotation rate \cite{muni06}.  This is in good agreement with the estimate based on the total quantized vorticity of the lattice.  For that we note the fact that in the rigid body limit $\Omega= (h N_v)/(2 m \pi R_{\rho}^2)$ \cite{nozi99}, and therefore, by measuring the number of vortices $N_v$ one can calculate the rotation rate.  We used $R = 37 \mu$m and $N_v$ was determined by a manual counting of the number of vortices from several images taken at long TOF. This resulted in $N_V = 37 \pm 7$, which leads to $\Omega=2\pi \times( 13.3 \pm 2.6$ Hz$) $.

\begin{figure}[htbp]\begin{center}
\includegraphics [width = \textwidth]{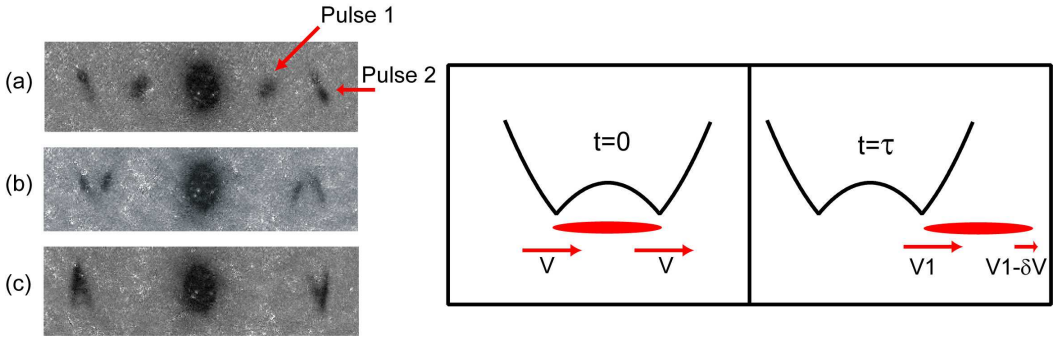}
\end{center}
\caption{In-Situ probe of the rotation.  Two consecutive pulses separated by a hold time $\tau$ can be used to non-destructively probe the rotating cloud.  Shown are 10 ms TOF images after $\tau = 7$ ms (a), $5$ ms (b), and $2.5$ ms (c).  The two diffracted groups appear as two tilted stripes on either side of the central, undiffracted cloud.  Motion within the trap causes the first group of atoms (pulse 1) to change its tilt angle due to kinematic considerations, as explained in the text.}\label{fig:tilting}
\end{figure}

\subsection{Non-Destructive Probing of the Rotation}

The time-of-flight technique is a destructive method and cannot be used for {\em in-situ} observation of the dynamics of the rotating condensate.  However, the Bragg method couples out a small fraction of the atoms, and therefore, can be used as a non-destructive probe.  We demonstrate this by applying a sequence of two $0.5$ ms duration Bragg pulses.  Pulse 1 diffracted a group of atoms which then evolved for a variable time delay $\tau$ in the harmonic trapping potential, after which a second group was created by pulse 2.  The trap was then immediately switched off, and an absorption image taken after 10 ms of time-of-flight.  The results are shown in Figure \ref{fig:tilting} for 3 different hold times $\tau$.  One can observe two diffracted stripes on either side of the undiffracted cloud in the center.  These correspond to the two groups of atoms created by pulses 1 and 2.  The stripes tilted to the left are the result of pulse 2, and resemble those observed in Figure \ref{Figure4: Spatial-Structures}b.  This indicates that the Bragg process is indeed non-destructive, and does not appear to significantly affect the rotation.  Due to the confining potential, atoms are slowed down during the time $\tau$ as they leave the cloud.  Therefore, the diffracted cloud from pulse 1 has a lower velocity than that of pulse 2 when the trap is shut off.  Therefore, it appears closer to the undiffracted cloud after a long TOF than does pulse 2.  Both pulses result in tilted images of the diffracted atoms; however, one can clearly see that pulse 1 has a tilt angle and shape which is different from pulse 2.  In fact, the tilt angle reverses sign after very short hold times $\tau > 2.5$ ms compared with the trap oscillation period of $\tau_{osc} = 32$ms.

This rapid distortion of the diffracted cloud from pulse 1 within 2.5 ms evolution in the trap is somewhat counterintuitive.  One might only expect the atoms to reverse their trajectory after $\tau_{osc}/4 = 8$ ms, that is, after a quarter period oscillation in the magnetic trap.  This apparent paradox can be resolved when one considers the spatial extent of the diffracted cloud and simple kinematics of the evolution.  In Figure \ref{fig:tilting} we show the full potential $V$ experienced by the Bragg scattered atoms.  This consists of both the harmonic trapping potential and the factor of 2 larger mean-field repulsion from the undiffracted condensate at density $n_c$, $V = V_{trap}+ 2 g n_c$ \cite{blak01}.  Both atoms at the front end of the BEC as well as those at the back receive a kick and begin to move with a recoil velocity $q/M$ which is considerably larger than the rotational velocity.  Atoms at the front ``climb up'' the potential for a time $\tau = 2.5$ms, and are therefore, slowed down, but atoms at the back end of the BEC do not change their energy substantially.  The back end can thus catch up with the front end during the 10 ms TOF.  Combined with the rotation of the strip, this causes a reversal of the tilt angle of the diffracted cloud from pulse 2.  We have performed a numerical simulation of classical trajectories in a two-dimensional harmonic potential for our parameters.  It confirms that the stripe should become vertical and tilt in the opposite direction for $\tau>2 $ms, which is consistent with our observations in Figure \ref{fig:tilting}c.  Therefore, the non-destructive method clearly has limitations, since the spatial TOF profile is the result of both the {\em initial} rotating distribution as well as its {\em in-situ} evolution during the hold time.

\section{Conclusion}

In conclusion, we have used Bragg scattering to directly measure the velocity profile of a rotating BEC.  The technique is complementary to time-of-flight imaging, and has direct application to the study of non-equilibrium superfluid dynamics \cite{donn91}, where the tangle of vortices reduces their visibility.  It might also be applied to measuring the normal fluid regime at finite temperature, where vorticity is not present.  Future experimental work will attempt to measure the microscopic flow field using Bragg diffraction.

We thank Andrew Seltzman for experimental assistance.  This work was supported by the U.S. Dept.\ of Energy, the Army Research Office and by Georgia Tech.

\bibliographystyle{elsart-num}

\begin{thebibliography}{10}
\expandafter\ifx\csname url\endcsname\relax
  \def\url#1{\texttt{#1}}\fi
\expandafter\ifx\csname urlprefix\endcsname\relax\def\urlprefix{URL }\fi

\bibitem{donn91}
R.~J. Donnelly, Quantized Vortices in Helium II, Cambridge Studies in Low
  Temperature Physics, Cambridge University Press, 1991.

\bibitem{bare01}
C.~F. Barenghi, R.~J. Donnelly, W.~F. Vinen, Quantized Vortex Dynamics and
  Superfluid Turbulence, Springer, 2001.

\bibitem{nels01}
D.~R. Nelson, Defects and Geometry in Condensed Matter Physics, 1st Edition,
  Cambridge University Press, 2001.

\bibitem{madi00}
K.~W. Madison, F.~Chevy, W.~Wohlleben, J.~Dalibard, Vortex formation in a
  stirred bose-einstein condensate, Physical Review Letters 84 (2000) 806--809.

\bibitem{abos01latt}
J.~R. Abo-Shaeer, C.~Raman, J.~M. Vogels, W.~Ketterle, Observation of vortex
  lattices in bose-einstein condensates, Science 292 (2001) 476--479.

\bibitem{halj01}
P.~Haljan, I.~Coddington, P.~Engels, E.~Cornell, Driving bose-einstein
  condensate vorticity with a rotating normal cloud, Physical Review Letters 87
  (2001) 210403.

\bibitem{hodb01}
E.~Hodby, G.~Hechenblaikner, S.~A. Hopkins, O.~M. Marago, C.~J. Foot, Vortex
  nucleation in bose-einstein condensates in an oblate, purely magnetic
  potential, Physical Review Letters 88~(1) (2001) 010405.

\bibitem{kozu99bragg}
M.~Kozuma, L.~Deng, E.~W. Hagley, J.~Wen, R.~Lutwak, K.~Helmerson, S.~L.
  Rolston, W.~D. Phillips, Coherent splitting of bose-einstein condensed atoms
  with optically induced bragg diffraction, Physical Review Letters 82 (1999)
  871.

\bibitem{sten99brag}
J.~Stenger, S.~Inouye, A.~P. Chikkatur, D.~M. Stamper-Kurn, D.~E. Pritchard,
  W.~Ketterle, Bragg spectroscopy of a bose-einstein condensate, Physical
  Review Letters 82 (1999) 4569--4573.

\bibitem{sims00}
J.~E. Simsarian, J.~Denschlag, M.~Edwards, C.~W. Clark, L.~Deng, E.~W. Hagley,
  K.~Helmerson, S.~L. Rolston, W.~D. Phillips, Imaging the phase of an evolving
  bose-einstein condensate wave function, Physical Review Letters 85~(10)
  (2000) 2040--2043.

\bibitem{thei04}
M.~Theis, G.~Thalhammer, K.~Winkler, M.~Hellwig, G.~Ruff, R.~Grimm, J.~H.
  Denschlag, Tuning the scattering length with an optically induced feshbach
  resonance, Physical Review Letters 93~(12) (2004) Art. No. 123001.

\bibitem{katz04}
N.~Katz, R.~Ozeri, J.~Steinhauer, N.~Davidson, C.~Tozzo, F.~Dalfovo, High
  sensitivity phonon spectroscopy of bose-einstein condensates using
  matter-wave interference, Physical Review Letters 93~(22) (2004) Art. No.
  220403.

\bibitem{fett01review}
A.~L. Fetter, A.~A. Svidzinsky, Vortices in a trapped dilute bose-einstein
  condensate, Journal of Physics-Condensed Matter 13~(12) (2001) R135--R194.

\bibitem{kevr04}
P.~G. Kevrekidis, R.~Carretero-Gonzalez, D.~J. Frantzeskakis, I.~G. Kevrekidis,
  Vortices in bose-einstein condensates: Some recent developments, Modern
  Physics Letters B 18~(30) (2004) 1481--1505.

\bibitem{pita03book}
L.~Pitaevskii, S.~Stringari, Bose-Einstein condensation, International Series
  of Monographs on Physics, Clarendon Press, Oxford, 2003.

\bibitem{petr95}
W.~Petrich, M.~H. Anderson, J.~R. Ensher, E.~A. Cornell, A stable, tightly
  confining magnetic trap for evaporative cooling of neutral atoms, Physical
  Review Letters 74 (1995) 3352.

\bibitem{peth02book}
C.~J. Pethick, H.~Smith, Bose-Einstein Condensation in Dilute Gases, Cambridge
  University Press, Cambridge, 2002.

\bibitem{nozi99}
P.~Nozières, D.~Pines, The Theory of Quantum Liquids, Perseus Books,
  Cambridge, Massachusetts, 1999.

\bibitem{stam01bragrev}
D.~M. Stamper-Kurn, A.~P. Chikkatur, A.~Gorlitz, S.~Gupta, S.~Inouye,
  J.~Stenger, D.~E. Pritchard, W.~Ketterle, Probing bose-einstein condensates
  with optical bragg scattering, International Journal of Modern Physics B
  15~(10-11) (2001) 1621--1640.

\bibitem{blak02}
P.~B. Blakie, R.~J. Ballagh, C.~W. Gardiner, Theory of coherent bragg
  spectroscopy of a trapped bose-einstein condensate, Physical Review A 65~(3)
  (2002) 033602.

\bibitem{zamb00}
F.~Zambelli, L.~Pitaevskii, D.~M. Stamper-Kurn, S.~Stringari, Dynamic structure
  factor and momentum distribution of a trapped bose gas, Physical Review A 61
  (2000) 063608.

\bibitem{muni06}
S.~R. Muniz, D.~S. Naik, C.~Raman, Bragg spectroscopy of vortex lattices in
  bose-einstein condensates, Physical Review A (Atomic, Molecular, and Optical
  Physics) 73~(4) (2006) 041605.

\bibitem{blak01}
P.~B. Blakie, R.~J. Ballagh, Spatially selective bragg scattering: A signature
  for vortices in bose-einstein condensates, Physical Review Letters 86~(18)
  (2001) 3930--3933.

\bibitem{cornkett99var}
E.~A. Cornell, W.~Ketterle, Bose-einstein condensation in atomic gases, in:
  M.~Inguscio, S.~Stringari, C.~E. Wieman (Eds.), Bose-Einstein condensation in
  Atomic Gases, Proceedings of the International School of Physics Enrico
  Fermi, Course CXL, Amsterdam, 1999, pp. 15--66.

\bibitem{madi01}
K.~W. Madison, F.~Chevy, V.~Bretin, J.~Dalibard, Stationary states of a
  rotating bose-einstein condensate: Routes to vortex nucleation, Physical
  Review Letters 86 (2001) 4443.





\end{thebibliography}


\end{document}